\documentclass[aps,prb,floatfix,superscriptaddress,twocolumn,footinbib]{revtex4-2}
\usepackage{amssymb}
\usepackage{amsmath}
\usepackage{amsfonts}
\usepackage{bm}
\usepackage{graphicx}
\usepackage{booktabs}
\usepackage{balance}
\usepackage[dvipsnames]{xcolor}
\usepackage{calc}
\usepackage{natbib}
\usepackage{color}
\usepackage[colorlinks,linkcolor=blue,anchorcolor=blue,citecolor=blue,urlcolor=blue]{hyperref}
\usepackage{lipsum}

\begin{document}
\title{Interplay of Altermagnetic Order and Wilson Mass in the Dirac Equation:
Helical Edge States without Time-Reversal Symmetry}

\author{Yu-Hao Wan}
\thanks{These authors contributed equally to this work.}
\affiliation{International Center for Quantum Materials, School of Physics, Peking
University, Beijing 100871, China}

\author{Peng-Yi Liu}
\thanks{These authors contributed equally to this work.}
\affiliation{International Center for Quantum Materials, School of Physics, Peking
University, Beijing 100871, China}

\author{Qing-Feng Sun}
\email[Corresponding author: ]{sunqf@pku.edu.cn.}
\affiliation{International Center for Quantum Materials, School of Physics, Peking
University, Beijing 100871, China}
\affiliation{Hefei National Laboratory, Hefei 230088, China}

\begin{abstract}
We investigate topological phases in three-dimensional topological insulator
(3DTI) thin films interfaced with altermagnetic (AM) orders.
Starting from a modified Dirac equation, we elucidate the interplay between the Wilson mass, arising from lattice regularization, and the altermagnetic mass, and show how this interplay fundamentally alters the band topology and boundary modes.
In particular, we demonstrate that coupling a 3DTI thin film to AM order induces a topological phase transition: although the total Chern number remains zero across the transition, topological helical edge states emerge after the transition.
These helical edge states arise from opposite Chern numbers at different high-symmetry points, and are distinct from both the chiral edge states of the quantum anomalous Hall phase and the helical edge states of the conventional quantum spin Hall states.
The quantum transport simulations reveal robust, quantized nonlocal resistance plateaus associated with these helical edge states,
which persist even under strong potential and magnetic disorder.
Our results establish 3DTI/AM heterostructures as a feasible material platform for engineering and detecting helical topological edge transport without time-reversal symmetry, thus expanding the landscape of topological matter and providing new opportunities for quantum devices.
\end{abstract}
 \maketitle

\section{Introduction}

The interplay between topology and magnetism has become a central theme in condensed matter physics, offering both fundamental insights and technological possibilities \cite{bhattacharyya_recent_2021,wan_magnetization-induced_2024,miao1,miao2,nsr,naref1}.
Topological insulators (TIs), with their symmetry-protected gapless edge or surface states \cite{fu_topological_2007,qi_topological_2008,zhang_topological_2009,hasan_colloquium_2010}, have provided a rich playground for exploring novel quantum phases, such as the quantum anomalous Hall and quantum spin Hall effects \cite{chang_colloquium_2023,chang_experimental_2013,bernevig_quantum_2006,bliokh_quantum_2015,kane_z_2005,kane_quantum_2005,konig_quantum_2007}.
In particular, the introduction of ferromagnetic (FM) order into three-dimensional topological insulator (3DTI) thin films has been extensively studied, leading to the realization of quantum anomalous Hall states with chiral edge conduction \cite{chang_experimental_2013,deng_quantum_2020}.
This FM/TI platform and its associated phenomenology are now well established both theoretically and experimentally.

At the heart of understanding topological phases in TIs is the modified Dirac equation, which captures band inversion and the emergence of topologically nontrivial states\cite{shen_topological_2012}. Here, "modified" refers to the inclusion of not only the conventional Dirac mass term but also an additional momentum-dependent Wilson mass. The Wilson mass was originally introduced to resolve the so-called "fermion doubling problem," which arises when the Dirac equation is formulated on a lattice \cite{nielsen_absence_1981,nielsen_no-go_1981,semenoff_condensed-matter_1984}. The Wilson mass acts as a regularizing term in the Dirac spectrum \cite{ginsparg_remnant_1982,kogut_hamiltonian_1975,wilson_confinement_1974,zhou_two-dimensional_2017}, and fundamentally determines the topological character of the resulting band structure. By tuning the relative strength and sign of the Dirac and Wilson mass terms, one can control whether the system realizes a trivial or nontrivial insulator \cite{shen_topological_2012}.

Recently, a new class of magnetic order, altermagnetism, has been discovered \cite{krempasky_altermagnetic_2024,mazin_prediction_2021,smejkal_beyond_2022}.
Altermagnets are characterized by collinear, spatially alternating spin structures and possess unconventional symmetries that distinguish them from both conventional ferromagnets and antiferromagnets.
As a result, they have attracted widespread research interest, with studies exploring their connections to superconductivity \cite{smejkal_emerging_2022,ouassou_dc_2023,chakraborty_zero-field_2024,cheng_field-free_2024,banerjee_altermagnetic_2024,giil_superconductor-altermagnet_2024,sun_andreev_2023,cheng_orientation-dependent_2024,SC1,SC2,SC3,JE1,JE2,JE3,JE4}, topological phenomena \cite{li_creation_2024,fernandes_topological_2024,li_floating_2025,TI1,TI2,TI3}, magnetic multipoles \cite{bhowal_ferroically_2024,fernandes_topological_2024}, anomalous Hall effect \cite{smejkal_anomalous_2022,chen_anomalous_2025}, parity anomaly \cite{wan_altermagnetism-induced_2025}, and magneto-optical effect \cite{naref2}.
When TIs are coupled to altermagnetic order, a new type of mass term—an ``altermagnetic mass" —may be introduced into the Dirac equation.
This mass term, with a distinct momentum and symmetry structure, offers the possibility to modulate the topological properties of TIs in ways fundamentally different from FM order.
However, how the interplay between the Wilson mass and the altermagnetic mass shapes the band topology and edge states in realistic systems remains largely unexplored.

A key open question is: How does the coexistence of Wilson mass and altermagnetic mass in the Dirac equation affect the band topology and the nature of edge states?
Can this interplay be realized in realistic material platforms, and does it give rise to topological phases and boundary modes beyond those in conventional quantum anomalous Hall and quantum spin Hall systems?
Addressing these questions not only enriches the classification of topological matter but also provides new avenues for designing functional quantum devices based on magnetic symmetry.

In this work, we establish a theoretical framework to investigate the role of an altermagnetic mass term in the Dirac equation, with a particular focus on its interplay with the Wilson mass.
In Sec. \ref{II}, our numerical analysis shows that this interplay induces momentum-dependent band inversions, leading to a classification of topological phases based on the Chern numbers at different high-symmetry points in the Brillouin zone.

To explore nontrivial topological phenomena beyond this idealized model, we consider a realistic 3DTI thin film coupled to altermagnetic order in Sec. \ref{III}.
In this system, surface hybridization and the AM-induced exchange field drive a topological phase transition: although the total Chern number remains zero, opposite Chern numbers coming from different high-symmetry points enable the formation of helical edge states.
Notably, these helical edge states are protected not by the conventional time-reversal symmetry \cite{kane_z_2005,kane_quantum_2005}, but by the crystalline and magnetic symmetries of the underlying lattice and altermagnetic order.

Our results show that while FM order leads to the conventional quantum anomalous Hall  phase with a single chiral edge state \cite{yu_quantized_2010,chang_experimental_2013,deng_quantum_2020}, AM order in 3DTI thin films gives rise to a pair of helical edge states tied to high-symmetry points—states that are fundamentally distinct from those in ordinary quantum spin Hall insulators.
These novel edge modes are closely related to recent advances in high-symmetry-point topological classification \cite{wan_classification_2025}.
Moreover, in Sec. \ref{IV}, quantum transport simulations based on non-equilibrium Green's function methods demonstrate that these helical edge states give rise to robust, quantized nonlocal resistance plateaus that persist even under strong potential and magnetic disorders, providing clear evidence of their topological protection.

This findings not only reveal a new mechanism for realizing helical edge transport without time-reversal symmetry, but also highlight the versatility of 3DTI/AM heterostructures as a tunable platform for topological phenomena.
These paves the way for further exploration of exotic boundary states and novel functionalities in magnetic topological materials.

\section{Dirac equation with AM mass}\label{II}

We begin with a minimal model for a Chern insulator, described by a modified Dirac equation \cite{shen_topological_2012,qi_topological_2006}: $H(\bm{k})=\bm{d}(\bm{k})\cdot\bm{\sigma}$, with
\begin{equation}
\bm{d}(\bm{k})=[Ak_{x},Ak_{y},m-B(k_{x}^{2}+k_{y}^{2})].\label{eq:2}
\end{equation}

Here, $A$ is a material-dependent parameter related to the Fermi velocity, $\bm{k} = (k_x, k_y)$ denotes the crystal momentum in the two-dimensional Brillouin zone, and $\bm{\sigma} = (\sigma_x, \sigma_y, \sigma_z)$ represents the vector of Pauli matrices acting on the spin degrees of freedom.

The topological properties are governed by two mass terms: the Dirac
mass $m$ and the Wilson mass $B$. Specifically, when $mB>0$, the
system realizes a nontrivial Chern insulator phase; the opposite sign
corresponds to a trivial insulator.

The introduction of $d$-wave altermagnetism additionally adds an altermagnetic term to the
Hamiltonian \cite{smejkal_beyond_2022,ifmmode_checkselse_sfimejkal_giant_2022,yi_spin_2025}, $J(k_{y}^{2}-k_{x}^{2})\sigma_{z}$, where $J$ quantifies
the strength of the altermagnetic effect. Both the Wilson mass term
and the altermagnetic mass term are quadratic in momentum and couple
to $\sigma_{z}$. However,
there are important differences: the altermagnetic mass term is anisotropic,
with opposite signs along the $x$ and $y$ directions, in contrast
to the isotropic Wilson mass. In fact, this anisotropy leads to significant
consequences for the topological properties: while the Wilson mass
produces a circular band inversion surface (BIS) in momentum space,
the altermagnetic mass produces a hyperbolic BIS. The BIS—defined
as the set of points in momentum space where the mass term vanishes
(i.e., where $d_{z}=0$)—plays a central role in the topological characterization
of the system \cite{wan_classification_2025}. The distinct geometries of the BIS for the Wilson and
altermagnetic masses thus result in fundamentally different topological
behaviors.

In the continuum model, momentum $k$ ranges from $-\infty$ to $+\infty$.
For the isotropic Wilson mass term, the sign of the mass at infinity
is the same in all directions and is solely determined by the sign
of $B$. As a result, the Wilson mass yields a single point at infinity,
forming a closed manifold where the Chern number is well defined.
In contrast, the altermagnetic mass term is anisotropic, with opposite
signs along the $x$ and $y$ directions. This means that as one approaches
infinity along the $k_{x}$ and $k_{y}$ axes, the direction of the
$\bm{d}(\bm{k})$ vector at infinity differs between these
directions. Consequently, the mass term does not converge to a single
point at infinity, preventing the definition of a proper closed manifold
and thus the Chern number in the continuum model. However, in a lattice
system, the Brillouin zone (BZ) forms a closed manifold with the topology
of a torus ($T^{2}$), which restores the proper definition of the
Chern number. Therefore, we study the system on a square lattice,
described by the Hamiltonian: $
H(\bm{k})=\bm{d}_{\rm lat}(\bm{k})\cdot\bm{\sigma}$, with
\begin{equation}
\bm{d}_{\rm lat}(\bm{k})=[A\sin k_{x},A\sin k_{y},m+2J(\cos k_{x}-\cos k_{y})].\label{eq:d-vec}
\end{equation}
Here, the traditional Wilson mass $2B(\cos k_x +\cos k_y -2)$
in the square lattice model is replaced by the altermagnetic term as appropriate.

\begin{figure*}
\centerline{\includegraphics[width=2\columnwidth]{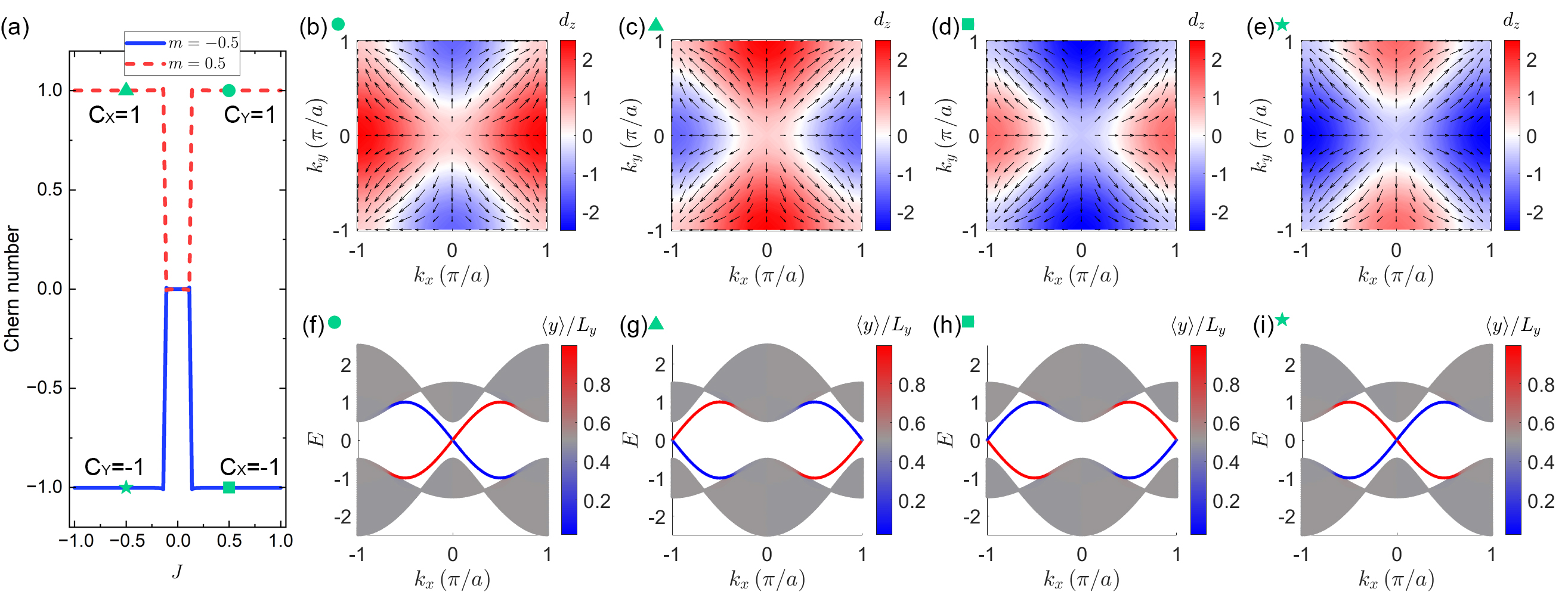}}
\caption{\label{fig:1}(a) Chern number \ensuremath{\mathcal{C}} as a function
of the altermagnetic strengthen $J$ for two representative Dirac
masses, $m>0$ (red) and $m<0$ (blue). The
four green markers highlight the parameter sets $J=\pm 0.5$ that are
analyzed in the remaining panels. (b-e) Spin-texture (d-vector) maps
in the full Brillouin zone for the four green points in (a). (f-i)
Band structures for nanoribbons of width $L_{y}$ = 50 corresponding
one-to-one to panels (b-e). The Wilson mass $B=0$.}
\end{figure*}

The Chern number can be calculated using \cite{shen_topological_2012}:
\begin{equation}
\mathcal{C}=\frac{1}{2\pi}\sum_{n}\int_{\mathrm{BZ}}d^{2}\bm{k}\;\Omega_{n}(\bm{k}),
\end{equation}
where the $\Omega_{n}(\bm{k})$ is momentum-dependent Berry curvature
for $n$-th band:
\begin{equation}
\Omega_{n}(\bm{k})=-\sum_{n'\neq n}\frac{2\,\mathrm{Im}\left[\langle\psi_{n\bm{k}}|v_{x}|\psi_{n'\bm{k}}\rangle\langle\psi_{n'\bm{k}}|v_{y}|\psi_{n\bm{k}}\rangle\right]}{(\varepsilon_{n'\bm{k}}-\varepsilon_{n\bm{k}})^{2}}
\end{equation}
with summation over all occupied bands. Here, $|\psi_{n\bm{k}}\rangle$
denotes the Bloch state of the $n$-th band, $v_{x}$, $v_{y}$ are
the velocity operators in the $x$ and $y$ directions, and $\epsilon_{n\bm{k}}$ is
the energy eigenvalue of the $n$-th band at momentum $\bm{k}$.

Fig. \ref{fig:1}(a) shows the evolution of the Chern number as a function
of the altermagnetic strength $J$. At $J=0$, the system is a trivial
insulator ($\mathcal{C}=0$). As $|J|$ increases, a topological phase
transition occurs, and the Chern number jumps to $+1$ (for $m>0$)
or $-1$ (for $m<0$). This indicates that the altermagnetic term
can drive the system into a topological phase.

Notably, for positive and negative values of $J$, the Chern number
can be the same. However, recent studies have shown that, in the presence
of inversion symmetry, topological phases with identical Chern numbers
can still be distinguished by the detailed configuration of skyrmions
in the Brillouin zone \cite{wan_classification_2025}.
In particular, the refined classification depends on which high-symmetry points are encircled by the BIS.
Our model preserves inversion symmetry, as indicated by $PH(\bm{k})P^{-1}=H(-\bm{k})$, with $P=\hat{P}\otimes\hat{R}_{2D}$, where $\hat{P}=\sigma_{z}$ acts on spin and $\hat{R}_{2D}$ denotes real-space inversion \cite{wan_classification_2025}. Consequently, the topology of the system can be classified according to the properties at these high-symmetry points.

Figs. \ref{fig:1}(b-e) illustrate the $\bm{d}(\bm{k})$ vector textures
in the Brillouin zone for representative parameter choices. The skyrmion charge (where $d_{x}=d_{y}=0$) are located at high-symmetry points in the Brillouin zone: $\Gamma(0,0)$, $X(\pi/a,0)$, $Y(0,\pi/a)$, and $M(\pi/a,\pi/a)$, with $a$ is the lattice constant.
In panels (b) and (c), the BIS [the white region in Figs. \ref{fig:1}(b-e)] encircles the $Y$ and $X$ points,
respectively. Although both cases correspond to a Chern number of
$+1$, the distinct skyrmion locations imply a finer topological distinction, which can be labeled as $\mathcal{C}_{Y}=1$ or $\mathcal{C}_{X}=1$, depending on the encircled high-symmetry point \cite{wan_classification_2025}.
Similarly, the phases shown in Figs. \ref{fig:1}(d,e) can be labeled as $\mathcal{C}_{X}=-1$ or $\mathcal{C}_{Y}=-1$, respectively.

This refined topological distinction is directly reflected in the
edge state spectra, as shown in Figs. \ref{fig:1}(f-i). For the phases with
$\mathcal{C}_{Y}=1$ and $\mathcal{C}_{X}=1$, edge states are localized near $k_{x}=0$ [Fig. \ref{fig:1}(f)] and $k_{x}=\pi$ [Fig. \ref{fig:1}(g)], respectively.
Thus, the momentum of the edge modes encodes the position of the corresponding skyrmion in the Brillouin zone.

Importantly, two phases with the same Chern number but different skyrmion
configurations cannot be adiabatically connected without closing the
bulk gap or breaking inversion symmetry. This necessitates the appearance
of gapless modes at their interface, highlighting the physical significance of this refined topological classification beyond the conventional Chern number \cite{wan_classification_2025}.

\begin{figure*}
 \centerline{\includegraphics[width=2\columnwidth]{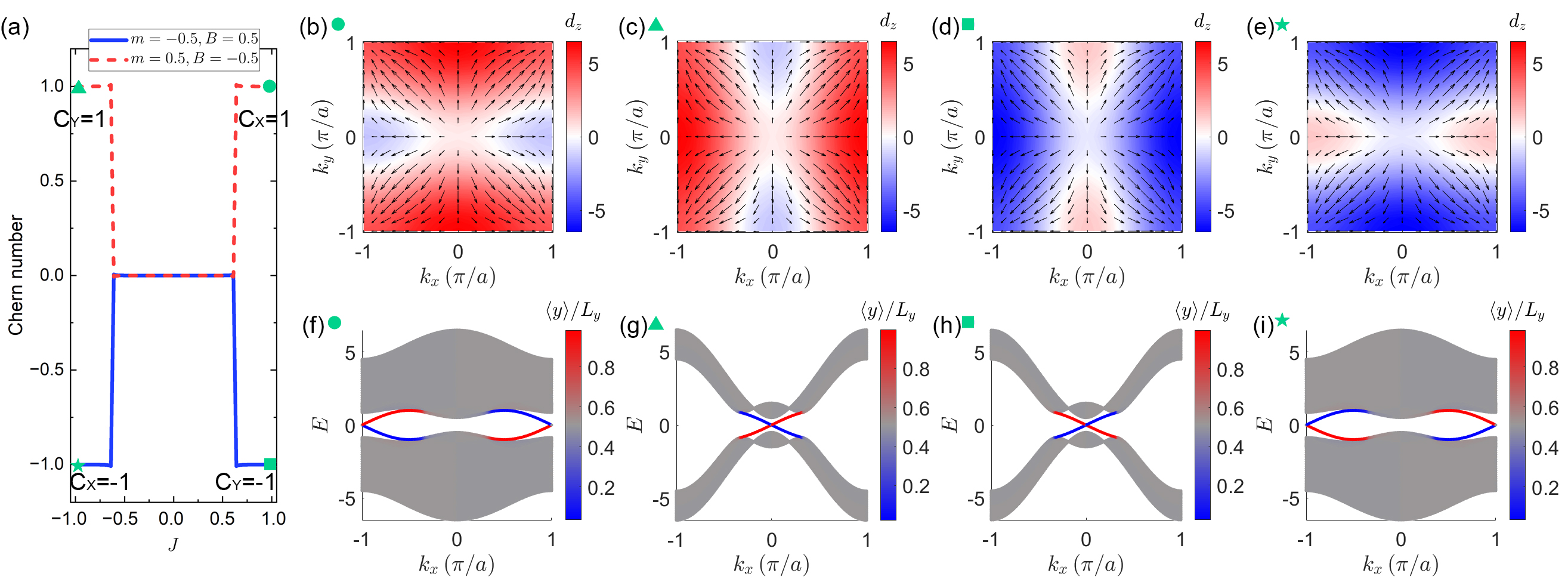}}
 \caption{\label{fig:2}(a) Chern number versus the altermagnetic parameter $J$ for $mB<0$.
 Four different shapes of green marks ($J =\pm1$) are indicated and they are analyzed in the remaining panels.
 (b-e) Corresponding skyrmion textures in the Brillouin zone.
 (f-i) Edge state spectra for nanoribbons with width $L_y=50$, at the same parameter
 values.}
\end{figure*}

As a short summary on Fig. \ref{fig:1}, we find that, on the lattice, the geometric difference between a circular BIS (arising
from the Wilson mass) and a hyperbolic BIS (from the altermagnetic
mass) leads to distinct Chern number classifications, governed by
which high-symmetry points are enclosed by the BIS. This provides
a route to engineer and distinguish topological phases with identical
Chern numbers but different underlying skyrmion structures.

Next, we study the coexistence of the altermagnetic mass and the Wilson mass.
Fig. \ref{fig:2}(a) shows the evolution of the Chern number as a function
of the altermagnetic mass $J$, with fixed parameters $mB=-0.5\times 0.5$
($m=\pm 0.5$, $B=\mp 0.5$).
For small $|J|$, the system is in a trivial phase ($\mathcal{C}=0$), dominated by the Wilson mass.
As $|J|$ increases, the altermagnetic mass overtakes the Wilson mass, and the system undergoes a topological
transition into a nontrivial phase with $\mathcal{C}=\pm1$, depending
on the sign of $m$.

This transition is illustrated in Figs. \ref{fig:2}(b-e), which show the BIS
and the corresponding $\bm{d}(\bm{k})$ textures for representative
parameter choices. Importantly, depending on the sign of $J$, the
BIS can encircle different high-symmetry points in the Brillouin zone—either
the $X$ or $Y$ point. For example, for $m=0.5$, with $J=1$, the BIS encloses
the $X$ point [Fig. \ref{fig:2}(b)]; with $J=-1$, it encloses the $Y$
point [Fig. \ref{fig:2}(c)]. The same logic applies for $m=-0.5$, as shown
in Figs. \ref{fig:2}(d-e), but with the Chern number sign reversed.

These distinctions manifest in the edge state spectra shown in Figs.
2(f-i). When the BIS encloses the $X$ point, $C_X=\pm 1$, and the chiral edge states appear near $k_{x}=\pi$ [Figs. \ref{fig:2}(f,i)].
When the BIS surrounds the $Y$ point, $C_Y=\pm 1$, and the edge states are localized near $k_{x}=0$ [Fig. \ref{fig:2}(g,h)].
Thus, while phases with the same Chern number can be realized, the location of their edge states and their physical properties
are distinct, directly tied to the BIS geometry.

In short, the sign of the Dirac mass term $m$ determines the sign
of the Chern number, and the sign of the altermagnetic mass $J$ selects
which high-symmetry point is encircled by the BIS (and thus the position
of the skyrmion in the Brillouin zone). This highlights the refined
topological classification enabled by the interplay between Wilson
and altermagnetic masses on the lattice system.

\begin{figure*}
 \centerline{\includegraphics[width=2\columnwidth]{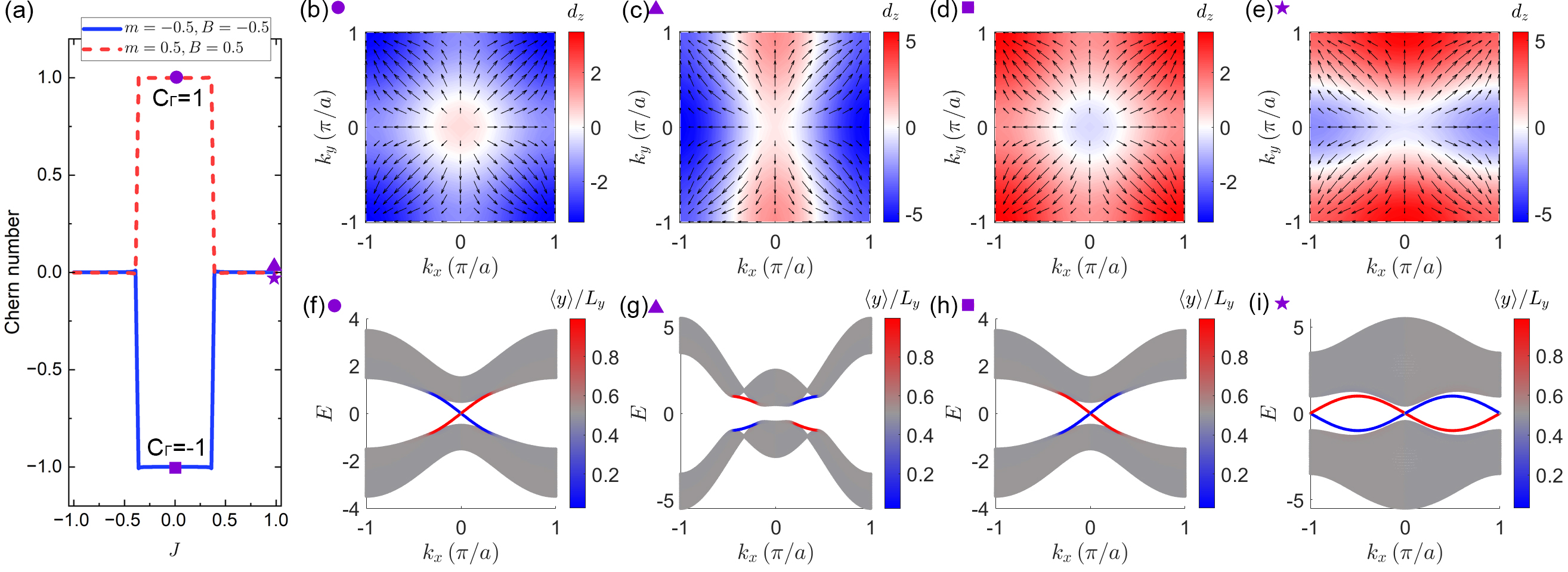}}
 \caption{\label{fig:3}(a) Chern number versus the altermagnetic parameter $J$ for $mB>0$.
 Four different shapes of purple marks ($J =\pm1$) are indicated and they are analyzed in the remaining panels. (b-e) Corresponding skyrmion textures in the Brillouin zone. (f-i) Edge state spectra for nanoribbons with
 width $L_y=50$, at the same parameter values.}
\end{figure*}

Fig. \ref{fig:3} investigates the evolution of topological phases when the
system is initially in a nontrivial regime with $mB>0$.
Fig. \ref{fig:3}(a) plots the
Chern number as a function of the altermagnetic mass $J$, with representative
points highlighted to illustrate the interplay between the Wilson
and altermagnetic masses.
As $J$ increases from $0$, the system gradually changes from a system with $\mathcal{C}=1$ to $\mathcal{C}=0$.

For $m=0.5$, $\ensuremath{B=0.5},$ and $J=0$, the BIS encloses
only the $\Gamma$ point [Fig. \ref{fig:3}(b)], resulting in a Chern number
$\mathcal{C}_\Gamma=1$. The corresponding nanoribbon edge spectrum [Fig. \ref{fig:3}(f)] clearly shows chiral edge states near $k_{x}=0$, characteristic
of a topological Chern insulator. As $J$ increases to 1, the BIS
evolves to simultaneously encircle both the $\ensuremath{\Gamma}$
and $Y$ points [Fig. \ref{fig:3}(c)]. In this configuration, the topological
charges at $\Gamma$ and $Y$ have opposite chiralities, causing their
contributions to the Chern number to cancel and driving the system
into a trivial phase with $\mathcal{C}=0$. The corresponding edge
spectrum [Fig. \ref{fig:3}(g)] shows that the edge states disappear from
the gap, confirming the loss of topological protection.

A similar evolution is observed for $m=-0.5$, $B=-0.5$. At $J=0$,
the BIS encloses the $\Gamma$ point [Fig. \ref{fig:3}(d)] with $\mathcal{C}_\Gamma=-1$,
and the edge spectrum [Fig. \ref{fig:3}(h)] shows chiral edge states near
$k_{x}=0$. However, when $J$ increases to 1, the BIS now encircles
both the $\Gamma$ and $X$ points [Fig. \ref{fig:3}(e)], again resulting
in a trivial total Chern number $\mathcal{C}=0$. The edge spectrum
[Fig. \ref{fig:3}(i)] displays in-gap states, but crucially, these edge-like
states are not topologically protected.

\begin{figure*}
 \centerline{\includegraphics[width=2\columnwidth]{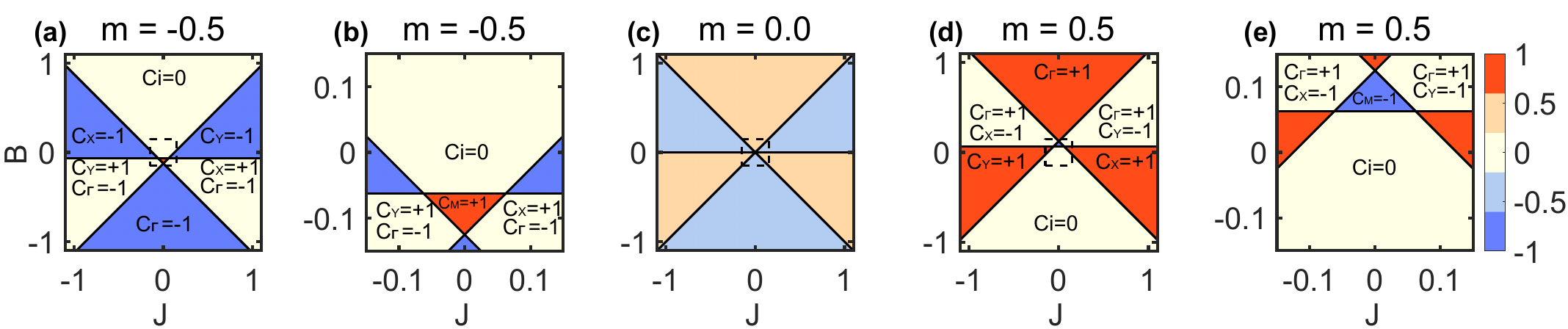}}
 \caption{\label{fig:pd}  The phase diagram of the Dirac equation with AM mass. (a), (c), and (d) are the cases where $m=-0.5$, $0$, and $0.5$ respectively. (b) and (e) are the enlarged graphs of the dashed boxes in (a) and (d) respectively. The colorbar represents the Chern number of each phase. Each phase is labelled by its Chern number with high-symmetry point, while $C_{\rm i}=0$ means that there is no band inversion.}
\end{figure*}

From the above discussion, we can see that the spin-texture and BIS of the Hamiltonian jointly determine the skyrmion structure of the system, which characterizes the topology.
Specifically in the model we discussed, BIS surrounding different high symmetry points led to different Chen number classifications.
So, we can identify all the phases of the Dirac equation with AM mass by calculating the band inversion situation of each high-symmetry point.
When the $z$ component of Eq. (\ref{eq:d-vec}) $d_{\rm lat}^{z}(\bm{k})$ is 0 at the high-symmetry points, the energy gap is closed, indicating the occurrence of a topological phase transition.
Therefore, the phase boundaries are given by Eq. (\ref{eq:pb}).
\begin{equation}
    \begin{aligned}
        &d_{\rm lat}^{z}(\Gamma)d_{\rm lat}^{z}(X)d_{\rm lat}^{z}(Y)d_{\rm lat}^{z}(M)\\
&=m(m-4B+4J)(m-4B-4J)(m-8B)=0
    \end{aligned}\label{eq:pb}
\end{equation}

According to Eq. (\ref{eq:pb}), we derive the phase diagram of the Dirac equation with AM mass, as shown in Fig. \ref{fig:pd}.
In Figs. \ref{fig:pd}(a,b), $m=-0.5$, phases with $\mathcal{C}_{X}=-1$, $\mathcal{C}_{Y}=-1$, $\mathcal{C}_{\Gamma}=-1$, and $\mathcal{C}_{M}=1$ emerge in the phase diagram, accompanied by trivial phases spaced between them.
Similarly, in Figs. \ref{fig:pd}(d,e), $m=0.5$, phases with $\mathcal{C}_{X}=+1$, $\mathcal{C}_{Y}=+1$, $\mathcal{C}_{\Gamma}=+1$, and $\mathcal{C}_{M}=-1$ emerge.
Interestingly, when $m=0$, as shown in Fig. {\ref{fig:pd}}(c), the system presents half-quantized Chen numbers due to its position at the topological phase transition point.
In summary, the altermagnetism brings rich topological classification to the Dirac equation in the lattice.
The edge states between different phases are well separated in the momentum space, suggesting potential edge state engineering.
Below, we demonstrated this possibility with a real system.

\section{helical edge state on 3DTI thin film with AM}\label{III}

\begin{figure*}
 \centerline{\includegraphics[width=2\columnwidth]{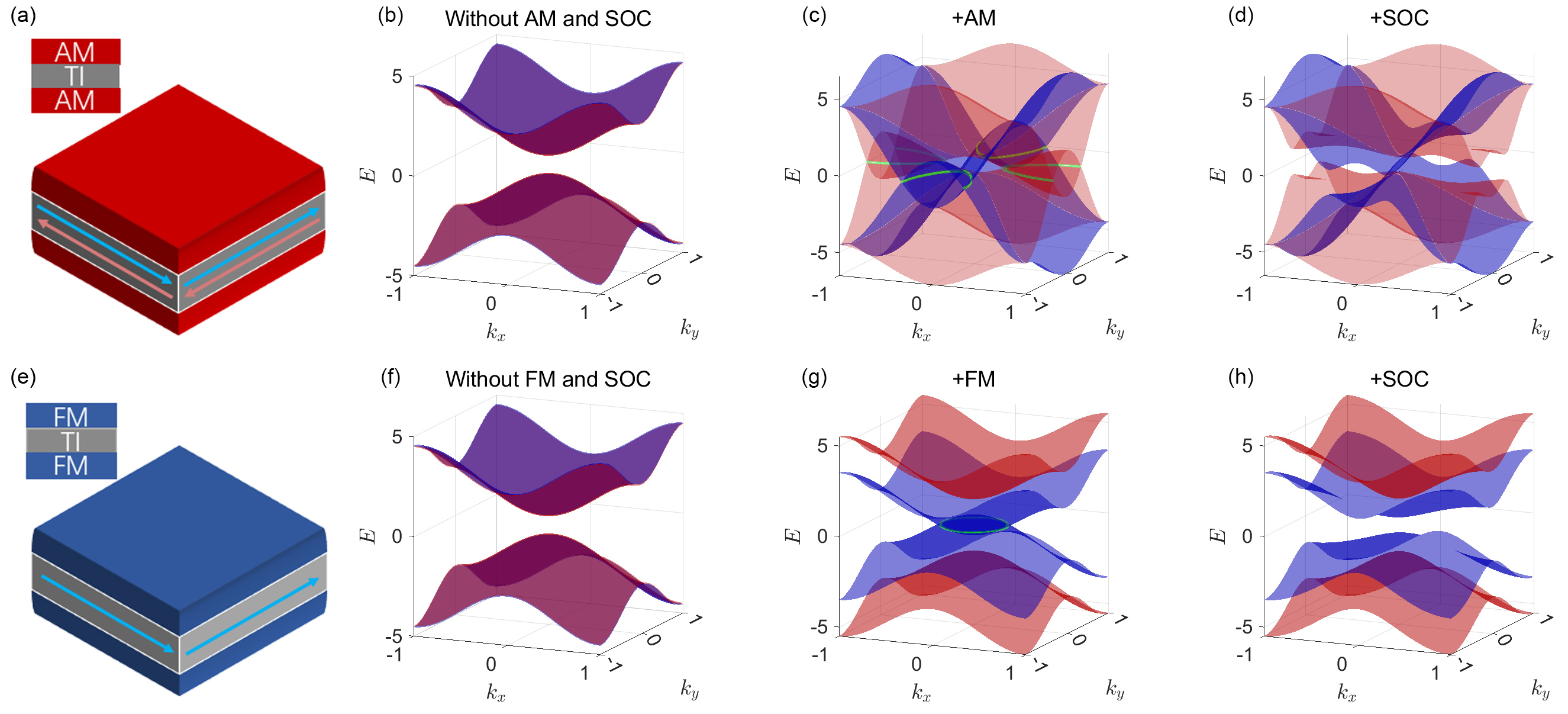}}
 \caption{\label{fig:4}(a,e) Schematics of 3DTI thin films (grey) with AM (dark red) and FM (dark blue) heterostructures, showing helical (light blue/red) and chiral (light blue) edge states, respectively.
 Band structures for 3DTI/AM (b-d) and 3DTI/FM (f-h) heterostructures under different conditions: without the magnetic order and SOC (b,f), with magnetic order but without SOC (c,g), and with magnetic order and SOC (d,h).
 Blue and red energy bands come from the two block-diagonal sectors;
 green lines denote the BIS.}
\end{figure*}

While the previous sections have established a thorough understanding
of the interplay between Wilson mass and altermagnetic mass in ideal
Dirac lattice models, such idealized Hamiltonians are primarily theoretical.
To bridge theory and experiment, we now propose a realistic platform
where these concepts can be realized: a thin film of a 3DTI with altermagnetic order, as summarized in Fig. \ref{fig:4}.
To show this, we start from the low-energy physics of a 3DTI thin film, which is governed by Dirac surface states localized on the top and bottom surfaces\cite{shen_topological_2012,yu_quantized_2010,lu_massive_2010}.
For sufficiently thin films, hybridization between these surface states opens a gap, and the system can be modeled by the following Hamiltonian in the basis $\left(|t\uparrow\rangle,|t\downarrow\rangle,|b\uparrow\rangle,|b\downarrow\rangle\right)$,
where $t/b$ denote the top/bottom surfaces and $\uparrow/\downarrow$ are for the $+z/-z$ spins:
\begin{equation}
H_{\mathrm{sf}}=\begin{bmatrix}0 & iv_{F}k_{-} & m_{k}^{*} & 0\\
-iv_{F}k_{+} & 0 & 0 & m_{k}^{*}\\
m_{k} & 0 & 0 & -iv_{F}k_{-}\\
0 & m_{k} & iv_{F}k_{+} & 0
\end{bmatrix},
\end{equation}
where $v_{F}$ is the Fermi velocity, $k_{\pm}=k_x \pm ik_y$, $k^{2}=k_{x}^{2}+k_{y}^{2}$, and $m_{k}=m_{0}-Bk^{2}$ describes the hybridization-induced mass.
 Hereafter, we set $v_{F}=1$ for convenience.

To explore the impact of different magnetic orders, we consider the
addition of either a ferromagnetic term $H_{\mathrm{FM}}=M\tau_{0}\otimes\sigma_{z}$ ($\tau$ represents the top/bottom degree of freedom) or an altermagnetic term $H_{\mathrm{AM}}$,
\begin{equation}
 \begin{aligned}
&H_{\mathrm{AM}}=\\
&\begin{bmatrix}J(k_{x}^{2}-k_{y}^{2}) & 0 & 0 & 0\\
0 & -J(k_{x}^{2}-k_{y}^{2}) & 0 & 0\\
0 & 0 & J(k_{x}^{2}-k_{y}^{2}) & 0\\
0 & 0 & 0 & -J(k_{x}^{2}-k_{y}^{2})
\end{bmatrix},
\end{aligned}
\end{equation}
where $M$ and $J$ represent the strengths of the FM and AM order, respectively.
 In this work, we focus on the case where the Néel vector of the altermagnetic order is aligned in the same direction on both the top and bottom surfaces. This setup is motivated by analogy to ferromagnetic TI heterostructures, where parallel exchange fields are believed to realize the QAH effect \cite{yu_quantized_2010}. Such a choice also allows for a clear comparison between the effects of ferromagnetic and altermagnetic proximity. Experimentally, recent studies have shown that the orientation of the N\'{e}el vector in altermagnetic materials can be controlled \cite{neel}, making it possible to realize such parallel configurations.

By transforming to the symmetric/antisymmetric basis,
\begin{equation}
 \begin{aligned}
|\pm \uparrow\rangle=\frac{|t\uparrow\rangle \pm |b\uparrow\rangle}{\sqrt{2}},
\hspace{5mm}
|\pm\downarrow\rangle=\frac{|t\downarrow\rangle \pm|b\downarrow\rangle}{\sqrt{2}},
\end{aligned}
\end{equation}
the Hamiltonian becomes block-diagonal under the basis: $\{|+\uparrow\rangle, |-\downarrow\rangle, |+\downarrow\rangle, |-\uparrow\rangle\}$ \cite{yu_quantized_2010},
\begin{equation}
H(\bm{k})=\begin{bmatrix}h_{\bm{k}}^+ & 0 \\
0 & h_{\bm{k}}^-
\end{bmatrix},
\end{equation}
the $\pm$ labels the two blocks, each corresponding to a different spin and surface parity.
Specifically, for the FM case,
\begin{equation}
h_{\bm{k}}^{(\pm)}=\left[m_{\bm{k}}\pm M\right]\sigma_{z}+v_{F}(k_{y}\sigma_{x}-k_{x}\sigma_{y}).
\end{equation}
And for the AM case,
\begin{equation}
h_{\bm{k}}^{(\pm)}=m_{\bm{k}}\sigma_{z}+v_{F}(k_{y}\sigma_{x}-k_{x}\sigma_{y})\pm J(k_{x}^{2}-k_{y}^{2})\sigma_{z}.
\end{equation}

To analyze the topology, we discretize the model on a square lattice, replacing \(k^{2}\) with \(2(1-\cos k_{x}) + 2(1-\cos k_{y})\), \(k_{x}^{2}-k_{y}^{2}\) with \(\cos k_{x} - \cos k_{y}\), and using \(\sin k_{x,y}\) for the linear momentum terms.
This procedure brings the Hamiltonian into the same form as Eq.(2), where the spin-orbit coupling (SOC) term
$v_F (\sin k_{y}\sigma_{x}-\sin k_{x}\sigma_{y})$
for the topological surface states corresponds to the \(A\sin k_{x}\sigma_x + A\sin k_{y}\sigma_y\) terms in Eq.(2).
The hybridization-induced mass term $m_{\bm{k}}\sigma_z$ in the thin film 3DTI model plays the same role as the Dirac mass and Wilson mass in the modified Dirac equation, while the momentum-dependent altermagnetic term $2J (\cos k_{x} - \cos k_{y})$
directly maps onto the altermagnetic mass term previously analyzed.
Thus, the altermagnetic 3DTI thin film model with block-diagonalized \(h^{\pm}\) is equivalent to the lattice version of the modified Dirac equation with opposite-sign altermagnetic mass terms, as discussed earlier. This establishes a direct correspondence between their parameters and topological features.
Since the hybridization-induced gap is topologically trivial, we set $m_0 =0.5$ and $B = -0.5$ in the following calculations.
We note that this lattice regularization may not capture all high-energy details at the Brillouin-zone boundaries. While higher-order terms away from the $\Gamma$ point are not included, the main topological results remain valid as long as the global bulk gap is open.

Fig. \ref{fig:4} presents a systematic comparison between the AM and FM scenarios.
For the AM case, Fig. \ref{fig:4}(a) illustrates the 3DTI/AM heterostructure,
where the red and blue lines indicate the helical edge states. The
calculated band structures in Fig. \ref{fig:4}(b-d) show that, without AM and
SOC, both blocks are trivially gaped [Fig. \ref{fig:4}(b)]. Turning
on AM order (but without SOC), both blocks acquire $k$-dependent
mass terms of opposite sign, producing BISs
at different high-symmetry points for each block (green lines in Fig. \ref{fig:4}(c)).
When SOC is included [Fig. \ref{fig:4}(d)], gaps open at both BISs, and
the two blocks acquire opposite Chern numbers ($\mathcal{C}_X=+1$ and $\mathcal{C}_Y=-1$), resulting in helical edge states.

In contrast, for the FM case, Fig. \ref{fig:4}(e) shows the schematic of a 3DTI thin film with a ferromagnetic heterostructure, where only a single
chiral edge state emerges \cite{yu_quantized_2010}.
The corresponding band structures [Figs. \ref{fig:4}(f-h)] reveal that, without FM and SOC, the system is again gapped [Fig. \ref{fig:4}(f)].
With FM order (but without SOC), the mass terms in the two blocks
shift in opposite directions, but only one block undergoes a topological
band inversion and hosts a circular BIS [green line in Fig. \ref{fig:4}(g)]. Upon including
SOC [Fig. \ref{fig:4}(h)], a gap opens at the BIS of the nontrivial block,
resulting in a single chiral edge mode characteristic of the quantum
anomalous Hall effect.

These results highlight the fundamental difference between AM and
FM orders in 3DTI thin films. While FM order yields a single chiral
edge state, AM order induces a pair of helical edge states, even in
the presence of magnetic ordering. The block-diagonalized effective
Hamiltonian provides a direct mapping from our ideal Dirac lattice
models to realistic material systems, confirming that key signatures
of symmetry-enriched topology, such as BISs, Chern numbers, and edge
modes, are all experimentally accessible.

To further substantiate the existence and robustness of the predicted
helical edge states in the AM case, we next perform quantum transport
calculations based on this model. These simulations will provide direct
evidence for the symmetry-enriched edge transport and support the
experimental feasibility of our proposal.

\section{robust nonlocal transport of helical edge states}\label{IV}

\begin{figure*}
 \centerline{\includegraphics[width=1.6\columnwidth]{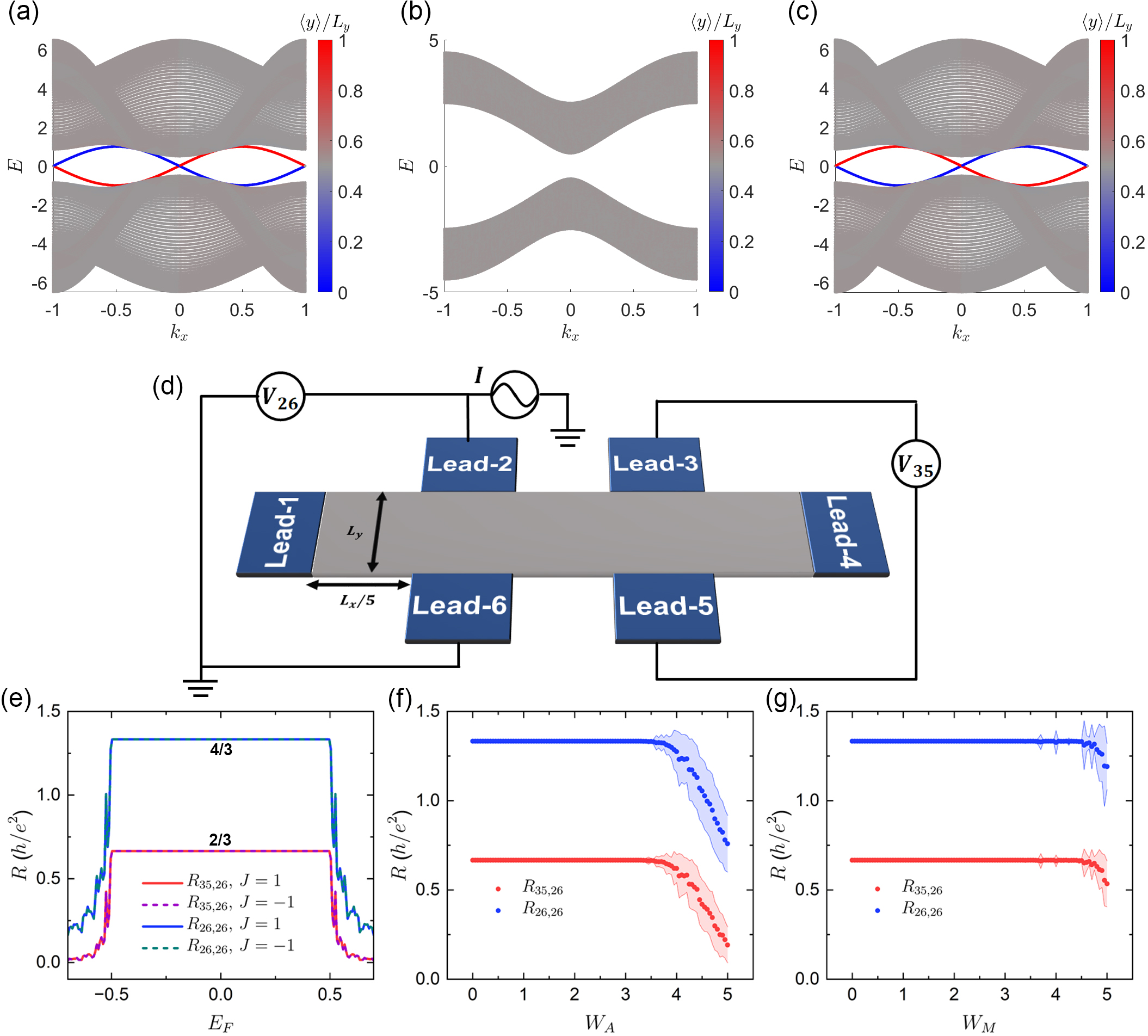}}
 \caption{\label{fig:5}Robust nonlocal transport of helical edge states.
 (a-c) show the band structure of a nanoribbon of 3DTI/AM film with $J=-1$, $0$, $1$, respectively.
 The colors of each Bloch state represent the centers of the corresponding wavefunctions $\langle y/L_y\rangle$, with $L_y=50$.
 (d) illustrates the setup of the nonlocal transport, where lead-2 and lead-6 serve as the source and drain, respectively.
 We set $L_y=L_x/5=50$.
 (e) is the numerical results of $R_{35,26}$ and $R_{26,26}$ for $J=\pm 1$, which quantize at $4h/3e^2$ and $2h/3e^2$ within the bulk gap, respectively.
 (f,g) demonstrate the robustness of quantized resistances against the potential energy disorder $W_A$ and magnetic disorder $W_M$.
 The red and blue dots are the average resistances $R_{35,26}$ and $R_{26,26}$ of 100 random configurations, respectively, with $J= 1$ and the Fermi energy $E_F=0$.
 The shaded regions show the corresponding standard deviations.}
\end{figure*}

To provide direct evidence for the existence of helical edge states
in the 3DTI/AM heterostructure, we first examine the electronic band
structure of a finite-width ribbon geometry.
Figs. \ref{fig:5}(a-c) display the calculated band spectra for different values of the altermagnetic mass parameter, $J=-1$, $0$, and $1$, respectively.
The color of each state encodes the center of the corresponding Bloch wavefunction $\langle y/L_{y}\rangle$, effectively distinguishing
whether a state is localized near the top or bottom edge of the sample.
For $J=0$, the system exhibits bulk bands without protected edge
modes.
In contrast, for $J=\pm 1$, one observes clearly separated
edge states traversing the bulk gap, with their spatial localization
swapping between the two boundaries as the momentum varies.
The presence of such counter-propagating is the hallmark of a helical topological phase, consistent with our theoretical prediction.

While the band structure analysis gives compelling theoretical evidence for edge states, it is essential to propose experimentally accessible transport signatures that can distinguish helical edge conduction from trivial or bulk-dominated transport.
In standard two-terminal conductance measurements, systems with helical edge states such as quantum spin Hall insulators typically yield a quantized
conductance $G$ close to $2e^{2}/h$, reflecting the contribution of two helical edge channels \cite{konig_quantum_2007}.
However, in small Hall bar geometries, such quantized conductance alone cannot unambiguously distinguish between ballistic bulk transport and genuine edge channel conduction, since both can yield similar values of $G$ in mesoscopic samples.
This ambiguity makes it challenging to conclusively identify the presence
of edge modes based solely on two-terminal measurements\cite{konig_quantum_2007}.

To unambiguously demonstrate the existence of edge channels, it is
therefore crucial to consider nonlocal transport measurements \cite{roth_nonlocal_2009}.
As shown schematically in Fig. \ref{fig:5}(d), we design a multi-terminal device geometry in which current is injected through one pair of leads
and the nonlocal voltage is detected at a spatially separated pair of edge leads.
In the presence of robust helical edge states, a finite nonlocal voltage signal is expected, as the edge channels allow selective transmission of carriers along the sample boundary, while the bulk remains insulating.
Such a nonlocal response serves as a definitive experimental signature of topological edge transport and is widely used to distinguish edge conduction from bulk effects in quantum spin Hall and related systems \cite{roth_nonlocal_2009,xing_nonlocal_2014}.
The low-energy transport of electrons is governed by the Landauer-B\"{u}ttiker formula \cite{datta_electronic_1995,jiang_topological_2009,wan_quarter-quantized_2024,liu_dissipation_2024},
\begin{equation}
I_p=\sum_{q}T_{pq}(V_p-V_q),\label{eq:LB}
\end{equation}
where $I_p$ and $V_p$ are the current and voltage of lead-$p$.
The transmission coefficient from lead-$q$ to lead-$p$ is $T_{pq}$.
When the transport of the system is completely controlled by the helical edge states, one can expect $T_{m,m+1}=T_{m+1,m}=1$ for $m=1$-$5$ and $T_{16}=T_{61}=1$.
The other matrix elements of the transmission coefficient matrix are all $0$ because the Fermi energy is located in the energy gap of the bulk \cite{roth_nonlocal_2009}.
Specifically, we respectively use lead-2 and lead-6 as the source and drain.
As shown in Fig. \ref{fig:5}(d), lead-2 is connected to the current source and lead-6 is grounded.
Other four leads act as voltage probes.
Substituting the transmission coefficients into Eq.(\ref{eq:LB}), one can obtain that in the ideal case of helical edge state transport, $[V_1,V_2,V_3,V_4,V_5,V_6]=[\frac{1}{2},1,\frac{3}{4},\frac{1}{2},\frac{1}{4},0]V_2$ and $I_2=-I_6=3e^2V_2/4h$.
Therefore, when measuring voltages $V_{26}=V_2-V_6$ and $V_{35}=V_3-V_5$, fractional quantized local resistance $R_{26,26}=V_{26}/I_2=4h/3e^2$ and non-local resistance $R_{35,26}=V_{35}/I_2=2h/3e^2$ should be obtained.
These measurement results will serve as clear evidence for helical edge states.

In the following, we will perform quantum transport simulations for
this nonlocal configuration, and demonstrate that the predicted symmetry-enriched helical edge states in the 3DTI/AM system indeed give rise to robust nonlocal transport signatures, thus paving the way for their experimental detection.
To quantitatively characterize the transport properties of the symmetry-enriched helical edge states, we perform non-equilibrium Green's function (NEGF) simulations for the multi-terminal geometry depicted in Fig. \ref{fig:5}(d).
In Eq.(\ref{eq:LB}), the transmission coefficient from lead-$q$ to lead-$p$ is given by $T_{pq}={\rm Tr}\left[ \mathbf{\Gamma}_p(E_F) \mathbf{G}^r(E_F) \mathbf{\Gamma}_q(E_F) \mathbf{G}^a(E_F) \right]$ \cite{datta_electronic_1995,meir_landauer_1992,jauho_time-dependent_1994}.
Here, $\mathbf{G}^r(E)=\left[\mathbf{G}^r(E)\right]^\dagger=\left[E-\mathbf{H}_{\rm cen}-\sum_p \mathbf{\Sigma}_p^r(E)+{\rm i}0^+\right]^{-1}$ is the retarded Green's function.
$\mathbf{H}_{\rm cen}$ is the discrete Hamiltonian of the center region [the grey region in Fig. \ref{fig:5}(d)] and $\mathbf{\Sigma}_p^r(E)$ is the retarded self-energy caused by the coupling of metal lead-$p$.
Without loss of generality, we set $\mathbf{\Sigma}_p^r(E)=-{\rm i}\mathbf{\Gamma}_p/2=-{\rm i}\mathbf{I}_p/2$, where $\mathbf{I}_p$ is a unit matrix.
Substituting the transmission coefficients calculated by the NEGF into Eq.(\ref{eq:LB}), one can obtain the real simulated current and voltage.

Specifically, we focus on the nonlocal resistance $R_{35,26}$ and
local resistance $R_{26,26}$ as functions of the Fermi energy for
both $J=-1$ and $J=1$, as shown in Fig. \ref{fig:5}(e).
The results with $J=\pm1$ are coincide, protected by a combined symmetry: $(J\rightarrow -J)\circ e^{-{\rm i}\sigma_z \pi/2}\circ e^{-{\rm i}\tau_x \pi/2}$.
Within the bulk energy gap, our calculations reveal quantized plateaus with $R_{35,26}=2h/3e^{2}$ and $R_{26,26}=4h/3e^{2}$, in excellent agreement with the theoretical expectation for two counter-propagating helical edge channels in this device configuration.
The fractional quantized values of nonlocal and local resistances directly reflect the presence of dissipationless edge transport, distinct from diffusive or bulk conduction.

To further evaluate the robustness of the helical edge states against disorder, we examine the impact of both Anderson-type potential disorder ($W_{A}$) and magnetic disorder ($W_{M}$) on the quantized resistance values.
Specifically, we model the potential (magnetic) disorder by adding a random onsite potential energy (Zeeman term), $w_a \tau_{0}\otimes\sigma_0$ ($w_m\tau_{0}\otimes \sigma_z$), to each site in the central region \cite{liu_four-terminal_2024,jiang_numerical_2009}.
The disorder strength is characterized by $w_a$ uniformly distributed within $[-W_A/2, W_A/2]$ and $w_m$ within $[-W_M/2, W_M/2]$.
As illustrated in Figs.~\ref{fig:5}(f,g), both $R_{35,26}$ and $R_{26,26}$ remain quantized at $2h/3e^{2}$ and $4h/3e^{2}$, respectively, over a broad range of $W_{A}$ and $W_{M}$.
This quantization persists until the disorder becomes strong enough to either close the bulk gap or induce significant hybridization between edge channels, thereby demonstrating the topological protection of the helical edge states against both nonmagnetic and magnetic perturbations.
 Moreover, even if the magnetic disorder exhibits long-range spin correlations, the helical edge states can still be well-preserved, and 
both the resistances $R_{35,26}$ and $R_{26,26}$ remain quantized (see Appendix A).

These results provide compelling evidence that the helical edge states
in the 3DTI/AM heterostructure are not only theoretically well defined,
but also manifest as robust, experimentally accessible transport signatures.
The quantized nonlocal resistances under various disorder strengths
serve as a clear hallmark of symmetry-enriched topological protection,
establishing the 3DTI/AM platform as a promising candidate for realizing
and detecting novel magnetic topological phases in experiment.

To assess the experimental relevance of our predictions, we briefly discuss viable materials for realizing 3DTI/altermagnet heterostructures.
For the 3DTI layer, well-established compounds such as Bi$_2$Se$_3$, Bi$_2$Te$_3$, and (Bi,Sb)$_2$Te$_3$ are commonly synthesized by molecular beam epitaxy (MBE), offering robust and clean topological surface states \cite{mbe}.
As for the altermagnetic layer, several promising materials have been recently identified, including RuO$_2$ , MnTe , and CrSb \cite{alter1,alter2,alter3,neel}.
Notably, MnTe has a hexagonal structure that is compatible with the lattice of Bi$_2$Se$_3$ and Bi$_2$Te$_3$, which facilitates high-quality interface formation. These combinations provide an experimentally feasible platform for exploring the predicted proximity-induced topological phenomena in 3DTI/altermagnet heterostructures. 

\section{Summary and conclusion}

In summary, we have systematically investigated the interplay between
the Wilson mass and the altermagnetic mass in the Dirac equation,
and demonstrated that this interplay gives rise to novel topological
phases classified according to high-symmetry points. By mapping these
theoretical insights onto a realistic model of a 3DTI thin film in
proximity to AM order, we establish a direct and experimentally
accessible platform for exploring the resulting topological phenomena.

Our study reveals that, the introduction of AM order generates a pair
of helical edge states that are fundamentally distinct from those
in conventional quantum anomalous Hall and quantum spin Hall systems. These helical edge states are
protected by the unique symmetries of the altermagnetic order and
are closely linked to recent developments in high-symmetry-point topological
classification \cite{wan_classification_2025}. Importantly, quantum
transport calculations show that these states manifest as robust,
quantized nonlocal resistance plateaus, providing clear experimental
signatures that are resilient to both potential and magnetic disorders.

Our work not only extends the theoretical understanding of symmetry-enriched
topological matter, but also paves the way for experimental realization
of new topological phases in magnetic TI heterostructures. The 3DTI/AM
platform offers a promising venue for engineering and detecting unconventional
edge states, thus enriching the landscape of topological materials
and opening up opportunities for future quantum device applications
based on magnetic symmetry and crystalline protection.

\begin{acknowledgments}
This work was financially supported by
the National Key R and D Program of China (Grant No. 2024YFA1409002),
the National Natural Science Foundation of China (Grants No. 12374034 and Grants No. 124B2069),
the Innovation Program for Quantum Science and Technology (Grant No. 2021ZD0302403),
The computational resources are supported by the High-Performance Computing Platform of Peking University.
\end{acknowledgments}

\section*{DATA AVAILABILITY}
The data that support the findings of this article are openly available\cite{data}.

\begin{figure*}
	\includegraphics[width=2\columnwidth]{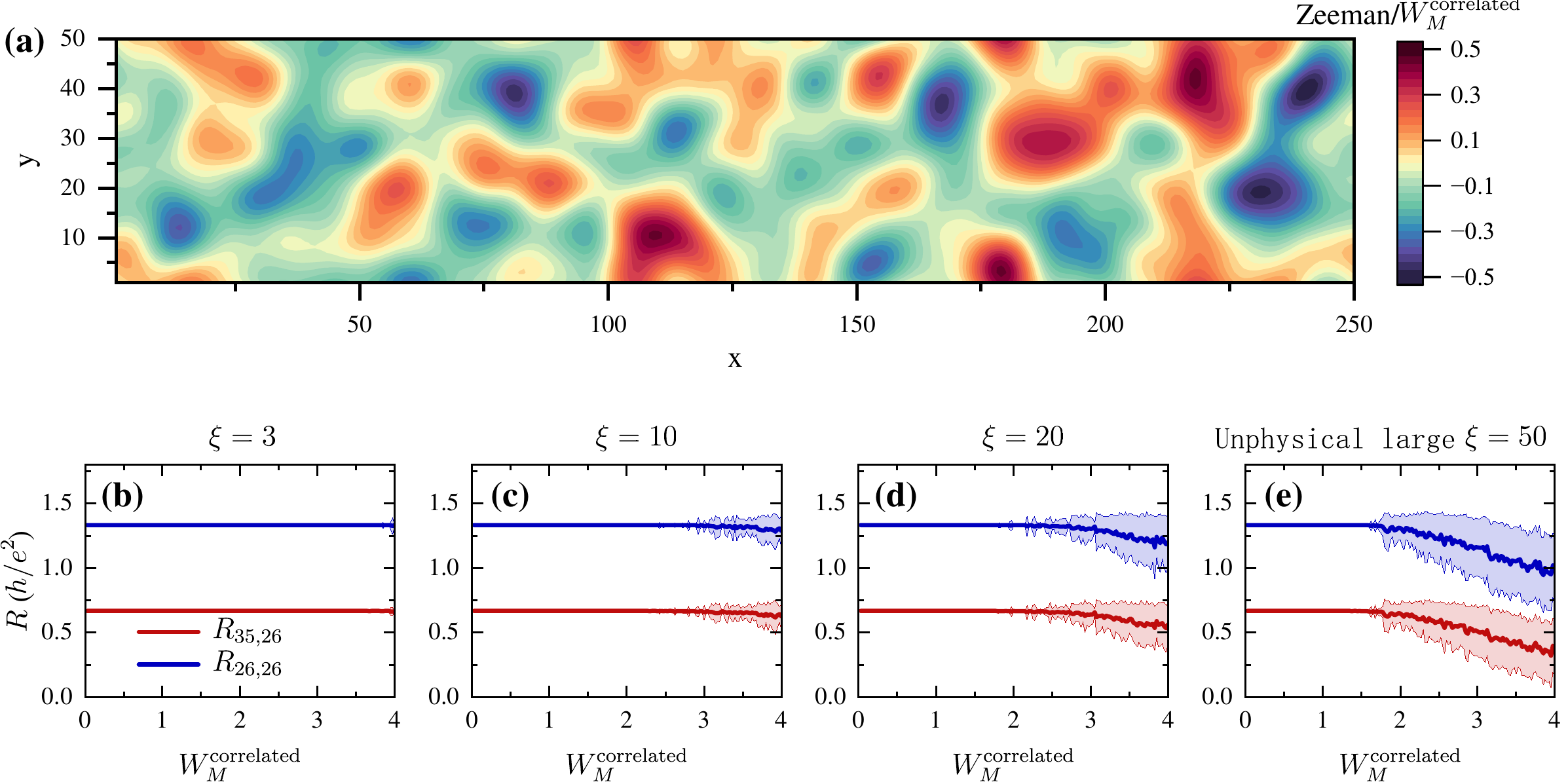}
			\caption{\label{fig:a}
				\baselineskip 4pt
{\color{black}
The effect of correlated magnetic disorder.
				(a) A typical real-space distribution of correlated magnetic disorder, generated by a completely random distribution within the range of $[-W_M^{\rm correlated}/2, W_M^{\rm correlated}/2]$ after Gaussian filtering.
                Correlation length $\xi=20$ is used here, in the same unit of $L_y$.
				(b-e) demonstrate the robustness of quantized resistances against the correlated magnetic disorder with different correlation length $\xi=3,10,20,50$, respectively.
                We choose $E_F = 0$ here, and each point is averaged over 100 random configurations.
                Other parameters used here are the same as those in Fig. \ref{fig:5}(b)}
				}
\end{figure*}

\appendix

\section{Correlated magnetic disorder}

In the main text, we treat the magnetic disorder as an uncorrelated random Zeeman field, as the discussion in Sec. \ref{IV}.
But in realistic systems, exchange (Heisenberg) interactions between local moments may give rise to spin correlations and magnetic domain formation.

To address this, we performed additional simulations incorporating the correlated (long-range) magnetic disorder. We use a Gaussian filter with correlation length $\xi$ (i.e., the average size of random magnetic domains) to smooth the fully randomed Zeeman field in the main text, which is a standard method to obtain a correlated disorder \cite{long1,long2,long3,wan_classification_2025}. As shown in Fig. \ref{fig:a}(a), this scenario faithfully models the correlated magnetic disorder observed in real magnetic microscope experiments \cite{long4}.

Importantly, we find that the helical transport remains robust against long-range magnetic disorder over a broad range of disorder strength $W_M^{\rm correlated}$ and correlation lengths $\xi$. As shown in Fig. \ref{fig:a}(b-e), the quantized resistances remains intact even when the disorder amplitude exceeds the full bulk gap ($W_M^{\rm correlated}>2m_0=1$).

For extremely large correlation lengths $\xi$ approaching the sample size (e.g., $\xi=50$ for our Hall bar of width $L_y=50$), the quantized resistences still maintain until the disorder amplitude reaches twice the bulk gap. Compared with small $\xi$, the robustness at this time is slightly weakened by the long-range magnetic order induced long-range backscattering. However, we emphasize that this represents an extreme and unphysical limit—essentially a globally coherent magnetic domain across the entire sample, which would correspond to complete destruction of any altermagnetic order.

Therefore, even when correlations in the magnetic disorder are taken into account, our results demonstrate that the system remains remarkably robust. Breakdown of helical transport occurs only when the disorder becomes extremely strong or when the underlying altermagnetic order is completely destroyed.

\bibliography{ref1}

\end{document}